\documentclass[%
preprint,
showpacs,preprintnumbers,
 amsmath,amssymb,
 aps,
 prc,
 longbibliography,
]{revtex4-1}

\usepackage{graphicx}% Include figure files
\usepackage{dcolumn}% Align table columns on decimal point
\usepackage{bm}% bold math
\usepackage{hyperref}% add hypertext capabilities

\begin{document}
%\setlength{\baselineskip}{20pt}

%\preprint{}

\title{Statistical theory of light nucleus reaction and application to $^9$Be(p, xn) reaction}


\author{Xiaojun SUN$^{1,3}$}
\email{sxj0212@gxnu.edu.cn}
\author{Jingshang ZHANG$^{2}$}

\affiliation{$^{1}$College of Physics, Guangxi Normal University, Guilin 541004, P. R. China}
\affiliation{$^{2}$China Institute of Atomic Energy, P. O. Box 275(41), Beijing 102413, P. R. China}
\affiliation{$^{3}$State Key Laboratory of Theoretical Physics, Institute of Theoretical Physics, Chinese Academy of Sciences, Beijing 100190, P. R. China}

%\thanks{}


\date{\today}

\begin{abstract}
A statistical theory of light nucleus reaction (STLN) is proposed to describe both neutron and light charged particle induced nuclear reactions with 1p-shell light nuclei involved. The dynamic of STLN is described by the unified Hauser-Feshbach and exciton model, of which the angular momentum and parity conservations are considered in equilibrium and pre-equilibrium processes. The Coulomb barriers of the incident and outgoing charged particles, which seriously influence the open reaction channels, could be reasonably considered in the incident channel and the different outgoing channels. In kinematics, the recoiling effects in various emission processes are taken strictly into account. Taking $^9$Be(p, xn) reaction as an example, we calculate the double-differential cross sections of outgoing neutrons and charged particles using PUNF code in the frame of STLN. The calculated results agree very well with the existing experimental neutron double-differential cross sections at $E_p=18$ MeV, and indicate that PUNF code is a powerful tool to set up file-6 in the reaction data library for the light charged particle induced nuclear reactions with 1p-shell light nuclei involved.

\end{abstract}


\maketitle

\section{\label{sect1}Introduction}

The 1p-shell light elements (Li, Be, B, C, N and O) had long been selected as the most important materials for improving neutron economy in thermal and fast fission reactors and in design of the accelerators driven spallation neutron sources, such as a candidate for target material in the intense neutron source of International Fusion Materials Irradiation Facility (IFMIF) \cite{Garin2011}, the plasma facing material of the first wall in International Thermonuclear Experimental Reactor (ITER) \cite{Marder2008}, the neutron multiplier in the fusion blanket \cite{Kawamura2002}, the neutron protection layer of Molten Salt Fast Reactor (MSFR) \cite{Yu2015}, the material of the accelerator based neutron source with a Fixed Field Alternating Gradient (FFAG) \cite{Mori2006} and the accelerator driven advanced nuclear energy system (ADANES) \cite{ZWL2015}. Additionally, some 1p-shell light elements are the materials in the determination of radiation shielding requirements for radiation protection purposes, optimization of dose delivery to a treatment volume, decisions on biological effectiveness of different therapy beams, and so on \cite{Sun2008kerma}. For these accurate designs of the target system, neutron shielding and nuclear medicine, the double-differential cross sections of the reaction products are very important as a source term for light particles (including neutron and charged particles) induced nuclear reactions with 1p-shell light nuclei involved.

For neutron induced nuclear reactions with 1p-shell light nuclei involved, the model calculations of the double-differential cross sections of reaction products have been successfully performed \cite{Zhang2015}. And file-6 has been established in CNDEL-3.1 library based on the theoretical calculation below 20 MeV incident energy \cite{Zhang2011}. File-6 is one of the most important files of nuclear reaction database and is recommended when the energy and angular distributions of the emitted particles must be coupled, when it is important to give a concurrent description of neutron scattering and particle emission, when so many reaction channels are open that it is difficult to provide separate reactions, or when accurate charged particle or residual nucleus distributions are required for particle transport, heat deposition, or radiation damage calculations \cite{Trkov2011}.

However, the double-differential cross sections of reaction products for light charged particle induced nuclear reactions with 1p-shell light nuclei involved are lack or not satisfactory till now, especially below 20 MeV incident energies. Taking p+$^9$Be reaction as an example, there are some coarse double-differential cross sections of reaction products only in ENDF/B-VII.1 \cite{Chadwick2011} and TENDL-2012 \cite{Koning2012}, and there are only some measured double-differential cross sections of outgoing neutrons at several incident energies. Several peaks of the measured double-differential cross sections of outgoing neutrons are observed for $^9$Be(p, xn) reaction. These peaks come mainly from the transitions between the discrete energy levels of the residual nuclei. The double-differential neutron-product cross sections of ENDF/B-VII.1, obtained by the Intranuclear-Cascade-Evaporation (ICE) model \cite{Young1990}, can not appropriately reproduce these experimental peaks (including positions and quantities), although ICE model are most applicable to a few hundreds MeV incident energies. The data of TENDL-2012, based on a software system built around the nuclear model code TALYS \cite{Koning2007}, can also not reasonably reproduce the double-differential cross sections for $^9$Be(p, xn) reaction. In addition, Monte Carlo calculation using Particle and Heavy Ion Transport System (PHITS) code is performed by the method combined the evaluated nuclear data files of ENDF/B-VII, the Bertini/GEM model and the JQMD/GEM model \cite{Iwamoto2009}. The calculated results can not also very well reproduce the experimental double-differential cross sections for $^9$Be(p, xn) reaction at 10 MeV incident energy. Recently, S. Hashimoto et al. proposed a new nuclear reaction model, which is a combination of the intranuclear cascade model and the distorted wave Born approximation, and used PHITS to estimate neutron spectra of reactions induced Li and Be by proton \cite{Hashimoto2014}. But there are some divergences between the calculated results and experimental double-differential cross sections for $^9$Be(p, xn) reaction at 39 MeV with 0 angle.

In addition, the continuum discretized coupled channels (CDCC) method is used to calculate the double-differential cross sections both for neutron and proton induced $^{6,7}$Li reactions \cite{Guo2013,Guo2014,Matsumoto2011,Ichinkhorloo2012}. Apart from the theoretical studies as far as I am concerned till now, there are no any publications of the double-differential cross sections for light charged particle induced reactions with the 1p-shell nuclei involved.

Although much effort has been made during the past several decades, there is lack of a general theory or method that can satisfactorily reproduce the measured double-differential cross sections for light charged particle induced nuclear reactions with the 1p-shell light nuclei involved. This problem may originate from several sources. Firstly, there is the absent theoretical method to describe the particle emission processes between the discrete levels of the residual nuclei with pre-equilibrium mechanism, which dominates all of the 1p-shell light nucleus reactions. Secondly, because of light mass, the recoil effect of the energy conservation must be strictly taken into account. Furthermore, there are individual features (including energy, spin, parity, width, branch ratio, and so on) of every energy level for each 1p-shell light nucleus. In this paper, the statistical theory of light nucleus reaction (STLN), which can describe the sequential and simultaneous particle emission processes between the discrete levels keeping conservations of energy, angular momentum and parity, is proposed to calculate the double-differential cross sections of outgoing neutrons and charged particles both for neutron and light charged particle induced reactions with the 1p-shell nuclei involved. Furthermore, taking $^9$Be(p, xn) reaction at 18 MeV as an example,
we calculate the double-differential cross sections of outgoing neutrons firstly using PUNF code in the frame of STLN. The calculated results agree very well with the existing experimental data.

This paper proceeds as follow. In Sec. \ref{sect2}, the dynamic and kinematics of STLN are introduced in detail. The reaction channels of p+$^9$Be reaction are analyzed detailedly, and the calculated results are compared with the experimental data in Sec. \ref{sect3}. In last section, a summary is given.

\section{Statistical theory of light nucleus reaction}\label{sect2}
It is assumed that the pre-equilibrium emission process from a compound nucleus to discrete levels
of the residual nuclei is the dominative reaction mechanism in light particle induced light nucleus reactions. Thus the dynamics of STLN can be described by the unified Hauser-Feshbach and exciton model \cite{Zhang1991,Zhang1993,Zhang1994}, which has applied successfully to calculate the double-differential cross sections of outgoing neutrons for neutron induced $^{6}$Li \cite{Zhang2001Li6}, $^{7}$Li \cite{Zhang2002Li7}, $^9$Be \cite{Sun2009Be9,Duan2010Be9}, $^{10}$B \cite{Zhang2003B10}, $^{11}$B \cite{Zhang2003B11}, $^{12}$C \cite{Zhang1999C12,Sun2007C12,Sun2008kerma}, $^{14}$N \cite{Yan2005N14}, $^{16}$O \cite{Zhang2001O16,Duan2005O16} and $^{19}$F \cite{Duan2007} reactions.

For conveniently describing the dynamics and kinematics, some quantities are defined as follows.

$M_T$: mass of the target nucleus with mass number $A_T$, proton number $Z_T$ and neutron number $N_T$;

$E_L$: kinetic energy of the incident particle in laboratory system;

$m_0$: mass of the incident particle with mass number $A_0$, proton number $Z_0$ and neutron number $N_0$;

$M_C$: mass of the compound nucleus with mass number $A_C=A_T+A_0$ and excited energy $E^*=\frac{M_T}{M_C}E_L+B_0$;

$m_1$ and $M_1$: masses of the first emitted particle and its residual nucleus, respectively;

$m_2$ and $M_2$: masses of the secondary particle emitted from $M_1$ and its residual nucleus, respectively;

$B_0, B_1$ and $B_2$: binding energies of $m_0$, $m_1$ in $M_C$ and $m_2$ in $M_1$, respectively;

$\varepsilon^X_{m_1}$ and $E^X_{M_1}$: kinetic energies of $m_1$ and $M_1$ in X coordinate system, respectively;

$\varepsilon^X_{m_2}$ and $E^X_{M_2}$: kinetic energies of $m_2$ and $M_2$ in X coordinate system, respectively;

Here, three motion systems will be used in STLN. Superscripts ($X=l, c, r$) denote the laboratory system (LS), the center-of-mass system (CMS) and the recoil nucleus system (RNS), respectively. For convenience, masses $m_i$ and $M_i$ ($i=0, C$, 1, 2) defined above also indicate the corresponding the nuclei or particles in text. It is obvious that there are approximate relations without lowering precision, i.e., $M_C\approx m_0+M_T\approx m_1+M_1$ and $M_1\approx m_2+M_2$.

\subsection{Dynamics}\label{sect2.1}
\subsubsection{First particle emission process}

In the frame of STLN, the cross section of the first emitted particle $m_1$ with kinetic energy $\varepsilon_{m_1}^c$ from compound nucleus $M_C$ to the $k_1$-th discrete energy level of residual nuclei $M_1$ can be described as
\begin{eqnarray}\label{eq2.1}
\sigma_{m_1,k_1}(E_L)=\sum_{j\pi}\sigma_a^{j\pi}(E_L)\{\sum_{n=3}^{n_{max}}P^{j\pi}(n)
\frac{W_{m_1,k_1}^{j\pi}(n,E^*,\varepsilon_{m_1}^c)}{W_T^{j\pi}(n,E^*)}
+Q^{j\pi}(n)\frac{W_{m_1,k_1}^{j\pi}(E^*,\varepsilon_{m_1}^c)}{W_T^{j\pi}(E^*)}\}.   \nonumber\\
\end{eqnarray}
Where, $P^{j\pi}(n)$ is the occupation probability of the $n$-th exciton state in the $j\pi$ channel ($j$ and $\pi$ denote the angular momentum and parity in final state, respectively), which
can be obtained by solving the $j$-dependent exciton master equation to conserve the angular momentum in pre-equilibrium reaction processes \cite{Zhang1994}. And $Q^{j\pi}(n)$ is the occupation probability of the equilibrium state in $j\pi$ channel expressed as
\begin{eqnarray}\label{eq2.2}
Q^{j\pi}(n)=1-\sum_{n=3}^{n_{max}}P^{j\pi}(n).
\end{eqnarray}

The absorption cross section $\sigma_a^{j\pi}(E_L)$ in $j\pi$ channel can be derived by Hauser-Feshbach statistical theory as \cite{Hauser1952}
\begin{eqnarray}\label{eq2.3}
\sigma_a^{j\pi}(E_L)=\frac{\pi}{k^2}\frac{(2j+1)}{(2I_T+1)(2s_0+1)}\sum_{S=|I_T-s_0|}^{I_T+s_0}
\sum_{l=|j-S|}^{min\{j+S,l_{max}\}}T_{l}(\varepsilon_{m_1}^c)g_l(\pi,\pi_T).
\end{eqnarray}
Where $I_T, \pi_T$ are the spin and parity of the target $M_T$, respectively. $s_0$ is the spin of incident particle $m_0$. And $k$ is the incident wave vector. $T_{l}(\varepsilon_{m_1}^c)$ is the reduced penetration factor of the first emitted particle $m_1$ \cite{Uhl1970} that can be obtained by the optical model of the spherical nucleus including the Coulomb barrier of incident charged particle.

In addition, parity conservation is determined by the orbit angular momentum $l$ of the relative motion between the incident particle $m_0$ and target nucleus $M_T$ in incident channel. For describing the parity conservation, we define the function
\begin{eqnarray}\label{eq2.4}
g_l(\pi,\pi_T)=\left\{
\begin{aligned}
1, ~~if ~\pi=(-1)^l \pi_T\\
0, ~~if ~\pi\neq(-1)^l \pi_T,
\end{aligned}
\right.
\end{eqnarray}
where $\pi$ and $\pi_T$ are the parities of the compound nucleus $M_C$ and the target nucleus $M_T$, respectively.

The emission rate $W_{m_1,k_1}^{j\pi}(n,E^*,\varepsilon_{m_1}^c)$ of the first emitted particle $m_1$ in Eq. (\ref{eq2.1}) at $n$-th exciton state with outgoing kinetic energy $\varepsilon_{m_1}^c$ can be expressed as
\begin{eqnarray}\label{eq2.5}
W_{m_1,k_1}^{j\pi}(n,E^*,\varepsilon_{m_1}^c) &=& \frac{1}{2\pi\hbar\omega^{j\pi}(n, E^*)}\sum_{S=|j_{k_1}-s_{m_1}|}^{j_{k_1}+s_{m_1}}\sum_{l=|j-S|}^{j+S}
T_{l}(\varepsilon_{m_1}^c)  \nonumber\\
&\times & g_l(\pi,\pi_{k_1})F_{m_1[\lambda,m]}(\varepsilon_{m_1}^c)Q_{m_1}(n)_{[\lambda,m]}.
\end{eqnarray}
Where, $\omega^{j\pi}(n, E^*)$ is the $n$-th exciton state density. $j_{k_1}$ is the angular momentum of the residual nucleus $M_1$ at energy level $E_{k_1}$, and $s_{m_1}$ is the spin of the first emitted particle $m_1$. $\pi$ and $\pi_{k_1}$ are the parities of the compound $M_C$ and residual nuclei $M_1$ at energy level $E_{k_1}$, respectively. The function of parity conservation are expressed as Eq. (\ref{eq2.4}) only substituting the target nucleus parity $\pi_T$ with the residual nucleus parity $\pi_{k_1}$.

In exciton model, $p$ and $h$ denote the particle number and hole number at $n$-th exciton state  ($n=p+h$), respectively. And $Q_{m_1}(n)_{[\lambda,m]}$, considering the effect of the incident particle memories, is the combination factor of the $n$-th exciton state expressed as\cite{Zhang1989}
\begin{eqnarray}\label{eq2.6}
Q_{m_1}(p,h)_{[\lambda,m]}&=& \left(\frac{A_T}{Z_T}\right)^{Z_{m_1}}\left(\frac{A_T}{N_T}\right)^{N_{m_1}}
\left(\begin{array}{c}p\\ \lambda\end{array}\right)^{-1}\left(\begin{array}{c}A_T-h\\ m\end{array}\right)^{-1}
\left(\begin{array}{c}A_{m_1}\\Z_{m_1}\end{array}\right)^{-1}       \nonumber\\
&\times& \sum_{i=0}^h \left(\begin{array}{c}h\\i\end{array}\right)\left(\frac{Z_T}{A_T}\right)^i\left(\frac{N_T}{A_T}\right)^{h-i} \nonumber\\
&\times& \sum_j \left(\begin{array}{c}Z_{m_1}+i\\j\end{array}\right) \left(\begin{array}{c}N_{m_1}+h-i\\\lambda-j\end{array}\right) \left(\begin{array}{c}Z-i\\Z_{m_1}-j\end{array}\right)
\left(\begin{array}{c}N-h+i\\N_{m_1}-\lambda+j\end{array}\right).
\end{eqnarray}
Where, $A_{m_1}, Z_{m_1}$ and $N_{m_1}$ are the mass number, proton number and neutron number of the first emitted particle $m_1$, respectively. And $A_{m_1}=\lambda+m$ denotes that there are $\lambda$ nucleon above the Fermi sea and $m$ nucleon blow the Fermi sea for the emitted particle $m_1$. The symbol $\left (\begin{aligned}n \\m\end{aligned}\right )$ is the binomial coefficient.

If the first emitted particle $m_1$ is nucleon, there is $\lambda=1, m=0$. Especially, if $m_1$ is neutron, i.e., $A_{m_1}=N_{m_1}=1$ and $Z_{m_1}=0$, thus Eq. (\ref{eq2.6}) can be simplified as \cite{Cline1972}
\begin{eqnarray}\label{eq2.7}
Q_n(p,h)_{[1,0]}=\left(\frac{A_T}{N_T}\right)\frac{1}{p}\sum_{i=0}^h \left(\begin{array}{c}h\\i\end{array}\right)\left(\frac{Z_T}{A_T}\right)^i\left(\frac{N_T}{A_T}\right)^{h-i}(N_{m_1}+h-i).
\end{eqnarray}
Similarly, if $m_1$ is proton, i.e., $A_{m_1}=Z_{m_1}=1$ and $N_{m_1}=0$, Eq. (\ref{eq2.6}) can be also simplified as \cite{Cline1972}
\begin{eqnarray}\label{eq2.8}
Q_p(p,h)_{[1,0]}=\left(\frac{A_T}{Z_T}\right)\frac{1}{p}\sum_{i=0}^h \left(\begin{array}{c}h\\i\end{array}\right)\left(\frac{Z_T}{A_T}\right)^i\left(\frac{N_T}{A_T}\right)^{h-i}(Z_{m_1}+i).
\end{eqnarray}
Obviously, if $m_1$ is $\gamma$ photo, i.e., $A_{m_1}=Z_{m_1}=N_{m_1}=0$, Eq. (\ref{eq2.6}) can be most simplified as $Q_{\gamma}(p,h)_{[0,0]}=1$. Apparently, the combination factor strictly keeps the particle conservation, i.e.,
\begin{eqnarray}\label{eq2.9}
\frac{N_T}{A_T}Q_n(p,h)_{[1,0]}+\frac{Z_T}{A_T}Q_p(p,h)_{[1,0]}=1.
\end{eqnarray}

In Eq. (\ref{eq2.5}), $F_{m_1[\lambda,m]}(\varepsilon_{m_1}^c)$ is the the pre-formation probability of composite particles $m_1$ at $n$-th exciton state in compound nucleus $M_C$, in which the momentum distributions of the exciton states are taken into account \cite{Zhang2007}. The consideration of the momentum distribution, which can improve Iwamoto-Harada model with no restriction in the momentum space, enhances the pre-formation probability of $[1,m]$ configuration and suppresses that of $[l>1,m]$ configurations seriously.

Considering the energy-momentum conservation in center of mass system (CMS), the definitive kinetic energies of the first emitted particle $m_1$ can be easily derived as
\begin{eqnarray}\label{eq2.10}
\varepsilon_{m_1}^c=\frac{M_1}{M_C}(E^*-B_1-E_{k_1}).
\end{eqnarray}
However, if the emitted particle $m_1$ is a charged particle, there is a threshold $E^{m_1}_{min}$ because of the Coulomb barrier $V_{Coul}\approx\frac{e^2Z_{m_1}Z_{M_1}}{R_c}$ ($R_c \approx r_c(A_{m_1}^{1/3}+A_{M_1}^{1/3}), r_c\approx 1.2\sim 1.5$ fm). Therefore, the open reaction channels must meet $E^*-E_{k_1}>B_1+V_{Coul}$. Obviously, Coulomb barrier can effect the open reaction channels seriously. And it is obvious that $T_{l}(\varepsilon_{m_1}^c)$ is 0, if $\varepsilon_{m_1}^c\leq E^{m_1}_{min}$.

Additionally, the total emission rate $W_T^{j\pi}(n,E^*)$ in pre-equilibrium reaction processes can be expressed as
\begin{eqnarray}\label{eq2.11}
W_T^{j\pi}(n,E^*)=\sum_{m_1,k_1}W_{m_1,k_1}^{j\pi}(n,E^*,\varepsilon_{m_1}^c).
\end{eqnarray}

In equilibrium reaction processes, the partial emission rate of the first particle $m_1$ in $j\pi$ channel and the total emission rate can be derived as following \cite{Hauser1952}
\begin{eqnarray}\label{eq2.12}
W_{m_1,k_1}^{j\pi}(E^*,\varepsilon_{m_1}^c)=\frac{1}{2\pi\hbar\rho^{j\pi}(E^*)}
\sum_{J=|j-I_{M_1}|}^{j+I_{M_1}}\sum_{l=|J-s_{m_1}|}^{J+s_{m_1}}
(2J+1)T_{Jl}(\varepsilon_{m_1}^c)g_l(\pi,\pi_{k_1}),
\end{eqnarray}
\begin{eqnarray}\label{eq2.13}
W_T^{j\pi}(E^*)=\sum_{m_1,k_1}W_{m_1,k_1}^{j\pi}(E^*,\varepsilon_{m_1}^c).
\end{eqnarray}
Where, $s_{m_1}$ and $I_{M_1}$ are the spins of the emitted particle $m_1$ and the corresponding residual nucleus $M_1$, respectively. $T_{Jl}(\varepsilon_{m_1}^c)$ is the penetration factor, and $\rho^{j\pi}(E^*)$ is the energy level density.

\subsubsection{Secondary particle emission process}

After the first particle $m_1$ emission, the cross section of the first residual nucleus $M_1$ at energy level $E_{k_1}$ emitting the secondary particle $m_2$ to the secondary residual nucleus $M_2$ at energy level $E_{k_2}$ can be expressed in the frame of STLN as follows
\begin{eqnarray}\label{eq2.14}
W_{m_2}^{j_{k_1}\pi_{k_1}\rightarrow j_{k_2}\pi_{k_2}}(E_{k_1}\rightarrow E_{k_2})=\frac{1}{2\pi}\sum_{S=|j_{k_2}-s_{m_2}|}^{j_{k_2}+s_{m_2}}\sum_{l=|j_{k_1}-S|}^{j_{k_1}+S}
T_l(\varepsilon_{m_2}^r)g_l(\pi_{k_1},\pi_{k_2}).
\end{eqnarray}
Where, $j_{k_1}\pi_{k_1}$ and $j_{k_2}\pi_{k_2}$ are the angular momentums and parities of the first and secondary residual nuclei, respectively. $T_{l}(\varepsilon_{m_2}^r)$ is the reduced penetration factor of the secondary emitted particle $m_2$, and $g_l(\pi_{k_1},\pi_{k_2})$ denotes the parity conservation in the secondary particle emission process.

The kinetic energy of the secondary emitted particle $m_2$ in recoil nucleus system (RNS) is expressed as
\begin{eqnarray}\label{eq2.15}
\varepsilon_{m_2}^r=\frac{M_2}{M_1}(E_{k_1}-B_2-E_{k_2}).
\end{eqnarray}

As well as $m_1$, there is a threshold $E^{m_2}_{min}$ if the secondary emitted particle $m_2$ is a charged particle because of the Coulomb barrier.

The total emission rate $W_T^{j_{k_1}\pi_{k_1}}(E_{k_1})$ from the first residual nucleus at energy level $E_{k_1}$ can be expressed as
\begin{eqnarray}\label{eq2.16}
W_T^{j_{k_1}\pi_{k_1}}(E_{k_1})=W_{\gamma}^{j_{k_1}\pi_{k_1}}(E_{k_1})+\sum_{m_2,k_2}
W_{m_2}^{j_{k_1}\pi_{k_1}\rightarrow j_{k_2}\pi_{k_2}}(E_{k_1}\rightarrow E_{k_2}).
\end{eqnarray}
Where, $W_{\gamma}^{j_{k_1}\pi_{k_1}}(E_{k_1})$ is the de-excited rate of $\gamma$ photo from energy level $E_{k_1}$.

So the branch ratio of the secondary emitted particle $m_2$ from energy level $E_{k_1}$ to the energy level $E_{k_2}$ can be expressed as
\begin{eqnarray}\label{eq2.17}
R_{m_2}^{k_1\rightarrow k_2}(E_{k_1})=\frac{W_{m_2}^{j_{k_1}\pi_{k_1}\rightarrow j_{k_2}\pi_{k_2}}(E_{k_1}\rightarrow E_{k_2})}{W_T^{j_{k_1}\pi_{k_1}}(E_{k_1})}.
\end{eqnarray}

Similarly, the branch ration of $\gamma$ photo can also be written as
\begin{eqnarray}\label{eq2.18}
R_{\gamma}^{k_1}(E_{k_1})=\frac{W_{\gamma}^{j_{k_1}\pi_{k_1}}(E_{k_1}}{W_T^{j_{k_1}\pi_{k_1}}(E_{k_1})}.
\end{eqnarray}

Thus, the cross section of the secondary particle $m_2$ emitting from the energy level $E_{k_1}$ to $E_{k_2}$ can be expressed as
\begin{eqnarray}\label{eq2.19}
\sigma_{k_1 \rightarrow k_2}(n, m_1, m_2)=\sigma_{k_1}(n, m_1)\cdot R_{m_2}^{k_1\rightarrow k_2}(E_{k_1}).
\end{eqnarray}

If the energy level $E_{k_1}$ is only de-excited by $\gamma$ photo to finish the reaction processes, the cross section of the first particle emission channel reads as
\begin{eqnarray}\label{eq2.20}
\sigma_{k_1}(n, m_1, \gamma)=\sigma_{k_1}(n, m_1)\cdot R_{\gamma}^{k_1}(E_{k_1}).
\end{eqnarray}

Eqs. (\ref{eq2.17})-(\ref{eq2.20}) describe the competitions among the reaction channels of the first particle emission, secondary particle emission and $\gamma$ de-excitation. In addition, the (reduced) penetration factor $T$ can be derived by optical model to fit the cross sections of all of the channels.

\subsection{Kinematics}\label{sect2.2}

\subsubsection{First particle emission process}

After the first particle $m_1$ is emitted, the residual nucleus $M_1$ is possible to remain at the energy level $E_{k_1}$. Considering the energy-momentum conservation in CMS, the definitive kinetic energies of $m_1$ and $M_1$ can be reexpressed for systematically describing the kinematics as
\begin{eqnarray}\label{eq3.1}
\varepsilon_{m_1}^c=\frac{M_1}{M_C}(E^*-B_1-E_{k_1})
\end{eqnarray}
and
\begin{eqnarray}\label{eq3.2}
E_{M_1}^c=\frac{m_1}{M_C}(E^*-B_1-E_{k_1}).
\end{eqnarray}

The normalized angular distributions of the first emitted particle $m_1$ and its residual nucleus $M_1$ with definitive kinetic energies can be standardized in nuclear reaction databases as \cite{Trkov2011}
\begin{eqnarray}\label{eq3.3}
\frac{d\sigma}{d\Omega^c_{Y}}=\sum_l\frac{2l+1}{4\pi}f_l^c(Y)P_l(\cos\theta^c_{Y}).
\end{eqnarray}
Here, $Y=m_1$ or $M_1$. $P_l(x)$ is the Legendre function, and the Legendre expansion coefficients $f^c_l(M_1)=(-1)^lf^c_l(m_1)$ can be derived from the the generilized master equation of the exciton model \cite{Zhang1989}.

Using the non-relativistic triangle relationship of the velocity vectors, the average kinetic energy of the first emitted particle $m_1$ in LS can be obtained
\begin{eqnarray}\label{eq3.4}
\overline{\varepsilon}^l_{m_1} &=& \int \frac{1}{2}m_1(\textbf{\textit{V}}_C+\textbf{\textit{v}}^c_{m_1})^2\frac{d\sigma}{d\Omega^c_{m_1}}d\Omega^c_{m_1}  \nonumber\\
 &=& \frac{m_1m_0E_L}{M^2_C}+\varepsilon^c_{m_1}+\frac{2}{M_C}\sqrt{m_0m_1E_L\varepsilon^c_{m_1}}f^c_1(m_1),
\end{eqnarray}
where $\textbf{\textit{V}}_C$ and $\textbf{\textit{v}}^c_{m_1}$ are the velocity vectors of the center-of-mass and the first emitted particle $m_1$ in CMS, respectively. Similarly as Eq. (\ref{eq3.4}), the average kinetic energy of the first residual nucleus $M_1$ in LS reads
\begin{eqnarray}\label{eq3.5}
\overline{E}^l_{M_1} = \frac{M_1m_0E_L}{M^2_C}+E^c_{M_1}-\frac{2M_1}{M_C}\sqrt{\frac{m_0E_LE^c_{M_1}}{M_1}}f^c_1(m_1).
\end{eqnarray}

Thus, it is obvious that the energy conservation for the first particle emission process in LS can be strictly kept as follows
\begin{eqnarray}\label{eq3.6}
E^l_{total}=\overline{\varepsilon}^l_{m_1}+ \overline{E}^l_{M_1}+E_{k_1}=E_L+B_0-B_1.
\end{eqnarray}

\subsubsection{Secondary particle emission processes}

For the 1p-shell light nucleus reactions, the secondary particle emission processes also come from the discrete energy levels after the first particle $m_1$ emission. There are four kinds of the particle emission processes as follows.

1) The residual nucleus $M_1$ at energy level $E_{k_1}$ emits the secondary particle $m_2$ with kinetic energy $\varepsilon^c_{m_2}$ to the secondary residual nucleus $M_2$ at energy level $E_{k_2}$.

2) The residual nucleus $M_1$ at energy level $E_{k_1}$ breaks spontaneously up into two particles.

3) The first emitted particle $m_1$ such as $^5$He, which is very unstable, breaks spontaneously up into a neutron and an alpha.

4) All of the first emitted particle $m_1$ and its residual nucleus $M_1$ is unstable and breaks spontaneously up into two smaller particles. This is so-called double two-body breakup reaction.

For the case 1), the residual nucleus $M_1$ at energy level $E_{k_1}$ with recoiling kinetic energy $E^c_{M_1}$ in CMS will emit the secondary particle $m_2$ with kinetic energy $\varepsilon^c_{m_2}$, if the conservations of the energy, angular momentum and parity are met. Thus, the corresponding residual nucleus $M_2$ at energy level $E_{k_2}$ will also gain the recoiling kinetic energy $E^c_{M_2}$ at arbitrary directions in CMS. In order to analytically describe the kinematics of the secondary emitted particle, we assume $M_1$ is static in RNS, then the definitive kinetic energy of the secondary emitted particle $m_2$ can be expressed as
\begin{eqnarray}\label{eq3.7}
\varepsilon^r_{m_2}=\frac{M_2}{M_1}(E_{k_1}-B_2-E_{k_2}).
\end{eqnarray}

Furthermore, the energy of the residual nucleus $M_2$ in RNS can be also obtained
\begin{eqnarray}\label{eq3.8}
E^r_{M_2}=\frac{m_2}{M_1}(E_{k_1}-B_2-E_{k_2}).
\end{eqnarray}

Using the non-relativistic triangle relationship $\textbf{\textit{v}}^c_{m_2}=\textbf{\textit{v}}^c_{M_1}+\textbf{\textit{v}}^r_{m_2}$, we can obtain \cite{Zhang2003B11,Zhang1999C12}
\begin{eqnarray}\label{eq3.9}
\varepsilon^c_{m_2}=\varepsilon^r_{m_2}(1+2\gamma\cos\Theta+\gamma^2),
\end{eqnarray}
\begin{eqnarray}\label{eq3.10}
\cos\Theta=\sqrt{\frac{\varepsilon^c_{m_2}}{\varepsilon^r_{m_2}}}[\cos\theta^c_{m_2}\cos\theta^c_{M_1}+\sin\theta^c_{m_2}\sin\theta^c_{M_1}\cos(\varphi^c_{m_2}-\varphi^c_{M_1})]-\gamma,
\end{eqnarray}
where $\gamma \equiv \sqrt{\frac{m_2E^c_{M_1}}{M_1\varepsilon^r_{m_2}}}$. The maximum and minimum kinetic energies of the secondary emitted particle $m_2$ in CMS are given by the following
\begin{eqnarray}\label{eq3.11}
\varepsilon^c_{m_2,max}=\varepsilon^r_{m_2}(1+\gamma)^2, ~~~~\varepsilon^c_{m_2,min}=\varepsilon^r_{m_2}(1-\gamma)^2.
\end{eqnarray}

In the frame of STLN, the double-differential cross section of the secondary emitted particle $m_2$ in RNS is assumed as the following isotropic distribution with a definitive kinetic energy $\varepsilon^r_{m_2}$, i.e.,
\begin{eqnarray}\label{eq3.12}
\frac{d^2\sigma}{d\varepsilon^r_{m_2}d\Omega^r_{m_2}}=\frac{1}{4\pi}\delta[\varepsilon^c_{m_2}-\varepsilon^r_{m_2}(1+2\gamma\cos\Theta+\gamma^2)].
\end{eqnarray}

Starting from the basic relation of the double-differential cross sections between CMS and RNS, the double-differential cross section of $m_2$ in CMS can be obtained through the corresponding results in RNS averaged by the angular distribution of the residual nucleus $M_1$, i.e.,
\begin{eqnarray}\label{eq3.13}
\frac{d^2\sigma}{d\varepsilon^c_{m_2}d\Omega^c_{m_2}}=\int d\Omega^c_{M_1} \frac{d\sigma}{d\Omega^c_{M_1}} \sqrt{\frac{\varepsilon^c_{m_2}}{\varepsilon^r_{m_2}}} \frac{d^2\sigma}{d\varepsilon^r_{m_2}d\Omega^r_{m_2}}.
\end{eqnarray}

By means of the properties of $\delta$ function and Eqs. (\ref{eq3.3})-(\ref{eq3.13}), the double-differential cross section of the secondary emitted particle $m_2$ in CMS can be rewritten as \cite{Zhang2015,Sun2015}
\begin{eqnarray}\label{eq3.14}
\frac{d^2\sigma}{d\varepsilon^c_{m_2}d\Omega^c_{m_2}}=\frac{1}{16\pi^2\gamma\varepsilon^r_{m_2}}\sum_l (2l+1)f_l^c(M_1)\int_0^{\pi} dtP_l(\sqrt{(1-\eta^2)\sin^2\theta^c_{m_2}}\cos t+\eta\cos\theta^c_{m_2}),         \nonumber\\
\end{eqnarray}
where $\eta=\sqrt{\frac{\varepsilon^r_{m_2}}{\varepsilon^c_{m_2}}}\frac{\varepsilon^c_{m_2}/
\varepsilon^r_{m_2}-1+\gamma^2}{2\gamma}$.
In term of the new integral formula \cite{Sun2015}, which has not been compiled in any integral tables or mathematical softwares, Eq. (\ref{eq3.14}) can be simplified as follows
\begin{eqnarray}\label{eq3.15}
\frac{d^2\sigma}{d\varepsilon^c_{m_2}d\Omega^c_{m_2}}=\sum_l \frac{(-1)^l}{16\pi\gamma\varepsilon^r_{m_2}}(2l+1)f_l^c(m_1)P_l(\eta)P_l(\cos\theta^c_{m_2}).
\end{eqnarray}

The normalized double-differential cross section of the secondary emitted particle $m_2$ is also standardized in nuclear reaction databases as \cite{Trkov2011}
\begin{eqnarray}\label{eq3.16}
\frac{d^2\sigma}{d\varepsilon^c_{m_2}d\Omega^c_{m_2}}=\sum_l\frac{2l+1}{4\pi}f_l^c(m_2)P_l(\cos\theta^c_{m_2}).
\end{eqnarray}

By comparing Eqs. (\ref{eq3.15}) and (\ref{eq3.16}), the Legendre expansion coefficients of the secondary emitted particle $m_2$ in CMS can be expressed as
 \begin{eqnarray}\label{eq3.17}
f_l^c(m_2)=\frac{(-1)^l}{4\gamma\varepsilon^r_{m_2}}f_l^c(m_1)P_l(\eta).
\end{eqnarray}

Similarly as Eq. (\ref{eq3.17}), we can also derive the analytical expression of the Legendre expansion coefficients of the secondary residual nucleus $M_2$ in CMS. The formula is expressed as \cite{Sun2015}
 \begin{eqnarray}\label{eq3.18}
f_l^c(M_2)=\frac{(-1)^l}{4\Gamma E^r_{M_2}}f_l^c(m_1)P_l(H),
\end{eqnarray}
where $\Gamma=\sqrt{\frac{M_2 E^c_{M_1}}{M_1E^r_{M_2}}}$ and $H=\sqrt{\frac{E^r_{M_2}}{E^c_{M_2}}}\frac{E^c_{M_2}/E^r_{M_2}-1+\Gamma^2}{2\Gamma}$.

It is obvious that the Legendre expansion coefficients of the secondary emitted particle $m_2$ and its residual nucleus $M_2$ in CMS are closely related to the first emitted particle $m_1$ and its recoiling nucleus $M_1$. Analytical expressions of Eq. (\ref{eq3.17}) and (\ref{eq3.18}) can largely reduce the volume of file-6 in nuclear reaction databases.

In CMS, the average kinetic energy of the secondary emitted particle $m_2$ can be obtained by averaging its double differential cross section, i.e.,
\begin{eqnarray}\label{eq3.19}
\overline{\varepsilon}^c_{m_2} &=& \int^{\varepsilon^c_{m_2,max}}_{\varepsilon^c_{m_2,min}}\varepsilon^c_{m_2}
\frac{d^2\sigma}{d\varepsilon^c_{m_2}d\Omega^c_{m_2}}d\varepsilon^c_{m_2}d\Omega^c_{m_2}  \nonumber\\
&=& \varepsilon^r_{m_2}(1+\gamma^2).
\end{eqnarray}

We also can obtain the average kinetic energy of the secondary residual nucleus $M_2$ in CMS in the same way, i.e.,
\begin{eqnarray}\label{eq3.20}
\overline{\varepsilon}^c_{M_2}= E^r_{M_2}(1+\Gamma^2).
\end{eqnarray}

In terms of the non-relativistic triangle relationship of the velocity vectors, the average kinetic energy of the secondary emitted particle $m_2$ in LS can be obtained
\begin{eqnarray}\label{eq3.21}
\overline{\varepsilon}^l_{m_2} &=& \int \frac{1}{2}m_2(\textbf{\textit{V}}_C+\textbf{\textit{v}}^c_{m_2})^2\frac{d^2\sigma} {d\varepsilon^c_{m_2}d\Omega^c_{m_2}}d\varepsilon^c_{m_2}d\Omega^c_{m_2}  \nonumber\\
 &=& \frac{m_0m_2E_L}{M^2_C}+\overline{\varepsilon}^c_{m_2}-2\frac{m_2}{M_C} \sqrt{\frac{m_0E_L E^c_{M_1}}{M_1}}f^c_1(m_1).
\end{eqnarray}

In the same way, the average kinetic energy of the secondary residual nucleus $M_2$ in LS can be derived as
\begin{eqnarray}\label{eq3.22}
\overline{E}^l_{M_2} = \frac{m_0M_2E_L}{M^2_C}+\overline{E}^c_{M_2}-2\frac{M_2}{M_C} \sqrt{\frac{m_0E_L E^c_{M_1}}{M_1}}f^c_1(m_1).
\end{eqnarray}

Thus, the energy conservation of the initial and final states for the light nucleus reactions can be strictly kept in LS as follows
\begin{eqnarray}\label{eq3.23}
E^l_{total} &=& \overline{\varepsilon}^l_{m_1}+\overline{\varepsilon}^l_{m_2}+ \overline{E}^l_{M_2}+E_{k_2} \nonumber\\
 &=& E_L+B_0-B_1-B_2.
\end{eqnarray}

For the case 2), the residual nucleus $M_1$ at energy level $E_{k_1}$ spontaneously break up into two smaller particles $m_2$ and $M_2$. It is assumed that $m_2$ and $M_2$ are at ground states, i.e., $E_{k_2}=0$. As well as the case 1), we assume $M_1$ is static in RNS, then the definitive kinetic energies of $m_2$ and $M_2$ can be expressed as
\begin{eqnarray}\label{eq3.24}
\varepsilon^r_{m_2}=\frac{M_2}{M_1}(E_{k_1}+Q_{M_2})
\end{eqnarray}
and
\begin{eqnarray}\label{eq3.25}
E^r_{M_2}=\frac{m_2}{M_1}(E_{k_1}+Q_{M_2}).
\end{eqnarray}
Where, $Q_{M_2}$ is the reaction $Q$ value for the breakup process $M_1\rightarrow m_2+M_2$.

Similarly, we can obtain the average kinetic energies of $m_2$ and $M_2$ in CMS as following
\begin{eqnarray}\label{eq3.26}
\overline{\varepsilon}^c_{m_2}=\frac{M_2}{M_1}(E_{k_1}+Q_{M_2})+\frac{m_1m_2}{M_1^2}\varepsilon^c_{m_1}
\end{eqnarray}
and
\begin{eqnarray}\label{eq3.27}
\overline{E}^c_{M_2}=\frac{m_2}{M_1}(E_{k_1}+Q_{M_2})+\frac{m_1M_2}{M_1^2}\varepsilon^c_{m_1}.
\end{eqnarray}

Furthermore, we can obtain the average kinetic energies of $m_2$ and $M_2$ in LS as following
\begin{eqnarray}\label{eq3.28}
\overline{\varepsilon}^l_{m_2}=\frac{m_0m_2E_L}{M_C^2}+\overline{\varepsilon}_{m_2}^c
-\frac{2m_2}{M_CM_1}\sqrt{m_0m_1E_L\varepsilon_{m_1}^c}f_1^c(m_1)
\end{eqnarray}
and
\begin{eqnarray}\label{eq3.29}
\overline{E}^l_{M_2}=\frac{m_0M_2E_L}{M_C^2}+\overline{E}_{M_2}^c
-\frac{2M_2}{M_CM_1}\sqrt{m_0m_1E_L\varepsilon_{m_1}^c}f_1^c(m_1)
\end{eqnarray}

Obviously, the energy conservation of the initial and final states can be strictly kept in LS as follows
\begin{eqnarray}\label{eq3.30}
E^l_{total} &=& \overline{\varepsilon}^l_{m_1}+\overline{\varepsilon}^l_{m_2}+ \overline{E}^l_{M_2} \nonumber\\
 &=& E_L+B_0-B_1+Q_{M_2}.
\end{eqnarray}

For the case 3), the first emitted particle $m_1$ can break spontaneously up into two smaller particles. For the 1p-shell light nucleus reactions, the unstable nucleus $m_1$ is only $^5$He, which breaks spontaneously up into a neutron ($m_n$) and an alpha ($M_{\alpha}$). As well as the case 1), we assume that $m_1$ is static in RNS. Then the definitive kinetic energies of $m_n$ and $M_{\alpha}$ can be expressed as
\begin{eqnarray}\label{eq3.31}
\varepsilon^r_{n}=\frac{M_{\alpha}}{m_1}Q_{m_1}
\end{eqnarray}
and
\begin{eqnarray}\label{eq3.32}
E^r_{\alpha}=\frac{m_n}{m_1}Q_{m_1}.
\end{eqnarray}
Where, $Q_{m_1}$ is the reaction $Q$ value for the breakup process $^5$He$\rightarrow$ n+$\alpha$.

Similarly, we can obtain the average kinetic energies of $m_n$ and $M_{\alpha}$ in CMS as following
\begin{eqnarray}\label{eq3.33}
\overline{\varepsilon}^c_{m_n}=\frac{M_{\alpha}}{m_1}Q_{m_1}+\frac{m_n}{m_1}\varepsilon^c_{m_1}
\end{eqnarray}
and
\begin{eqnarray}\label{eq3.34}
\overline{E}^c_{M_{\alpha}}=\frac{m_n}{m_1}Q_{m_1}+\frac{M_{\alpha}}{m_1}\varepsilon^c_{m_1}.
\end{eqnarray}

Furthermore, we can obtain the average kinetic energies of $m_n$ and $M_{\alpha}$ in LS as following
\begin{eqnarray}\label{eq3.35}
\overline{\varepsilon}^l_{m_n}=\frac{m_0m_nE_L}{M_C^2}+\overline{\varepsilon}_{m_n}^c
+\frac{2m_n}{M_Cm_1}\sqrt{m_0m_1E_L\varepsilon_{m_1}^c}f_1^c(m_1)
\end{eqnarray}
and
\begin{eqnarray}\label{eq3.36}
\overline{E}^l_{M_{\alpha}}=\frac{m_0M_{\alpha}E_L}{M_C^2}+\overline{E}_{M_{\alpha}}^c
+\frac{2M_{\alpha}}{M_Cm_1}\sqrt{m_0m_1E_L\varepsilon_{m_1}^c}f_1^c(m_1)
\end{eqnarray}

Obviously, the energy conservation of the initial and final states can be strictly kept in LS as follows
\begin{eqnarray}\label{eq3.37}
E^l_{total} &=& \overline{E}^l_{M_1}+\overline{\varepsilon}^l_{m_n}+ \overline{E}^l_{M_{\alpha}} \nonumber\\
 &=& E_L+B_0-B_1+Q_{m_1}.
\end{eqnarray}

It is worth mentioning that the symbols before the Legendre expansion coefficients $f_1^c(m_1)$ of $m_n$ and $M_{\alpha}$ are positive in case 3) comparing the negative signs in other cases. This is due to the forward tendency of the first emitted particle $m_1$.

For the case 4), the first emitted particle $m_1$ and its residual nucleus $M_1$ break spontaneously up into two smaller particles at the same time. As well as case 2), the averaged kinetic energies of $m_2$ and $M_2$ in both CMS and LS coming from the residual nucleus $M_1$ can be expressed as Eqs. (\ref{eq3.24})-(\ref{eq3.29}). For this double two-body breakup process of the 1p-shell light nucleus reactions, the first emitted particle $m_1$ is $^5$He, which can spontaneously break up into a neutron ($m_n$) and an alpha ($M_{\alpha}$). The averaged kinetic energies of neutron ($m_n$) and alpha ($M_{\alpha}$) in both CMS and LS coming from  $^5$He can be expressed as Eqs. (\ref{eq3.31})-(\ref{eq3.36}). Thus, the energy conservation of the initial and final states can be strictly kept in LS as follows
\begin{eqnarray}\label{eq3.38}
E^l_{total} &=& \overline{E}^l_{M_2}+\overline{\varepsilon}^l_{m_2}
+\overline{E}^l_{M_{\alpha}}+\overline{\varepsilon}^l_{m_n} \nonumber\\
 &=& E_L+B_0-B_1+Q_{m_1}+Q_{M_2}.
\end{eqnarray}

%
%\section{Angular distribution of the first emitted particle}
%
%For the 1p-shell light nucleus reactions, the first particle is always emitted from the compound nucleus $M_C$ at excited energy $E^*$ to its residual nucleus $M_1$ at discrete energy levels $E_{k_1}$.
%Because of the definitive kinetic energies in CMS expressed Eq. (\ref{eq3.1}) and (\ref{eq3.2}),
%so the first emitted particle $m_1$ and its residual nucleus $M_1$ can be described very well by using the angular distributions. The form of the angular distribution is recommended in the nuclear reaction database as Eq. (\ref{eq3.3}). So the Legendre expansion coefficients $f_l^c(m_1)$ in CMS is very important not only to describe the first emitted particles but also to express the double-differential cross sections of the secondary emitted particles as mentioned in Sec. \ref{sect2.2}.
%
%

\begin{figure}
\centering
\includegraphics[width=24cm,angle=0]{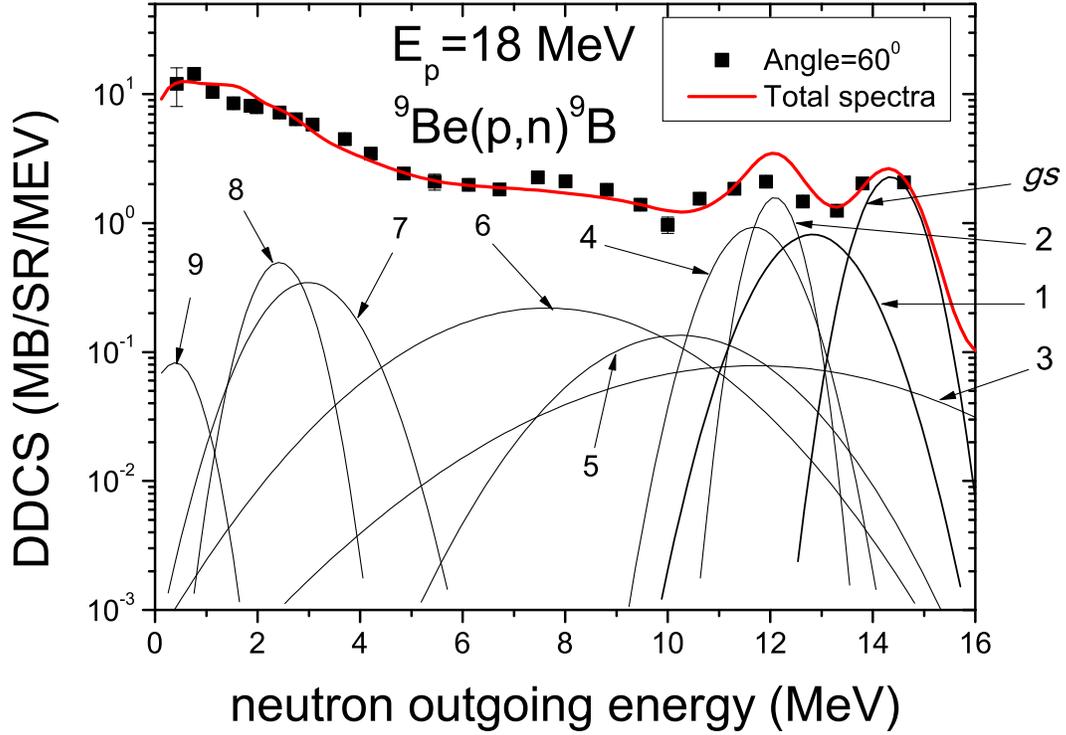}
\caption{(Color online) The partial double-differential cross sections of outgoing neutrons from reaction channel (p, n)$^9$B with outgoing angle 60$^0$ at $E_p$=18 MeV in LS. The points denote the experimental data taken from Ref. \cite{Verbinski1969}, and the red solid line denotes the calculated total double-differential cross sections. The black solid lines denote the partial spectra of the first emitted neutron from the compound nucleus to the ground state, up to 9th excited energy levels (as the numbers labeled in figure) of the first residual nucleus $^9$B, in which broadening effects must be taken into account. Only the cross sections with the values larger than 0.1 mb are given.  }\label{Fig1}
\end{figure}

\begin{figure}
\centering
\includegraphics[width=24cm,angle=0]{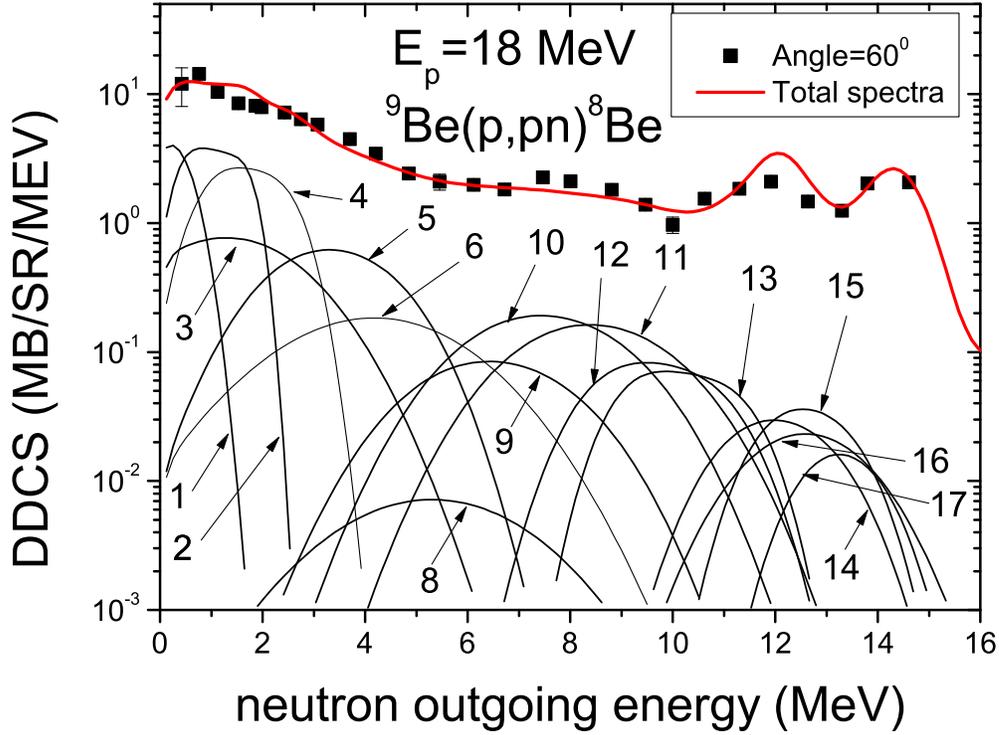}
\caption{(Color online) The same as Fig. \ref{Fig1}, but the partial double-differential cross sections from reaction channel (p, pn)$^8$Be$\rightarrow$(p, pn+2$\alpha$). The black solid lines denote the partial spectra of the secondary emitted neutron from the 1st-17th excited energy levels (as the numbers labeled in figure) of the first residual nucleus $^9$Be to the ground state of the secondary residual nucleus $^8$Be. }\label{Fig2}
\end{figure}

\begin{figure}
\centering
\includegraphics[width=24cm,angle=0]{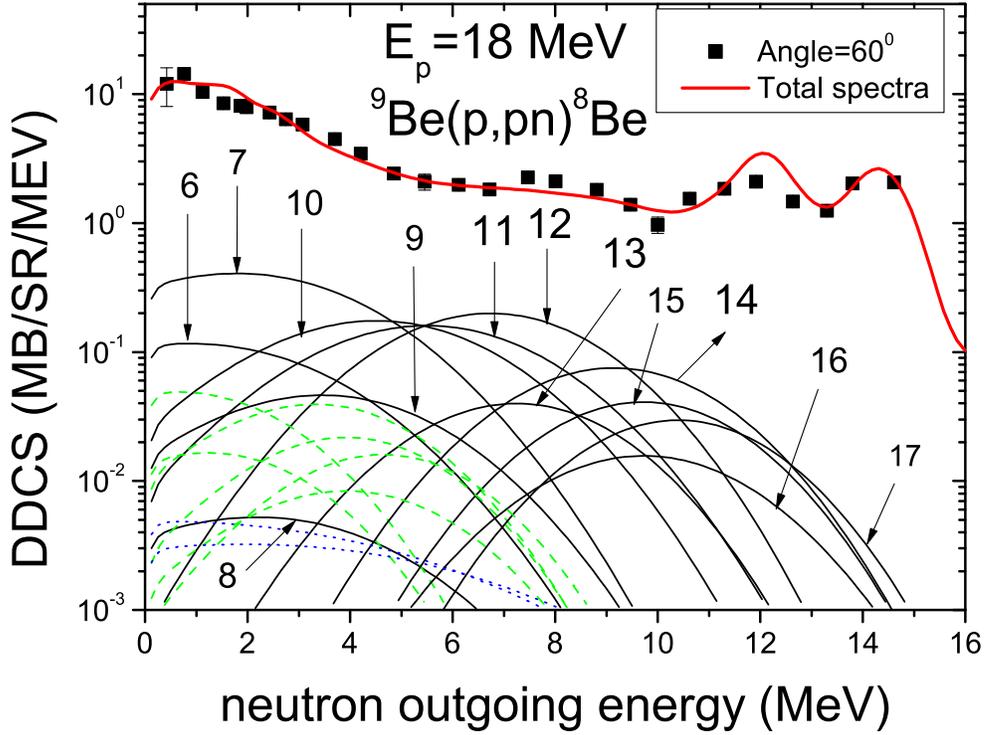}
\caption{(Color online) The same as Fig. \ref{Fig2}, but the black solid lines denote the partial spectra from the 6th-17th excited energy levels (as the numbers labeled in figure) of $^9$Be to the first excited energy level of $^8$Be. The green dash lines denote the partial spectra from the 12th-17th excited energy levels of $^9$Be to the second excited energy level of $^8$Be, and the blue dot lines denote the partial spectra from the 14th and 17th excited energy levels of $^9$Be to the third excited energy level of $^8$Be.}\label{Fig3}
\end{figure}

\begin{figure}
\centering
\includegraphics[width=24cm,angle=0]{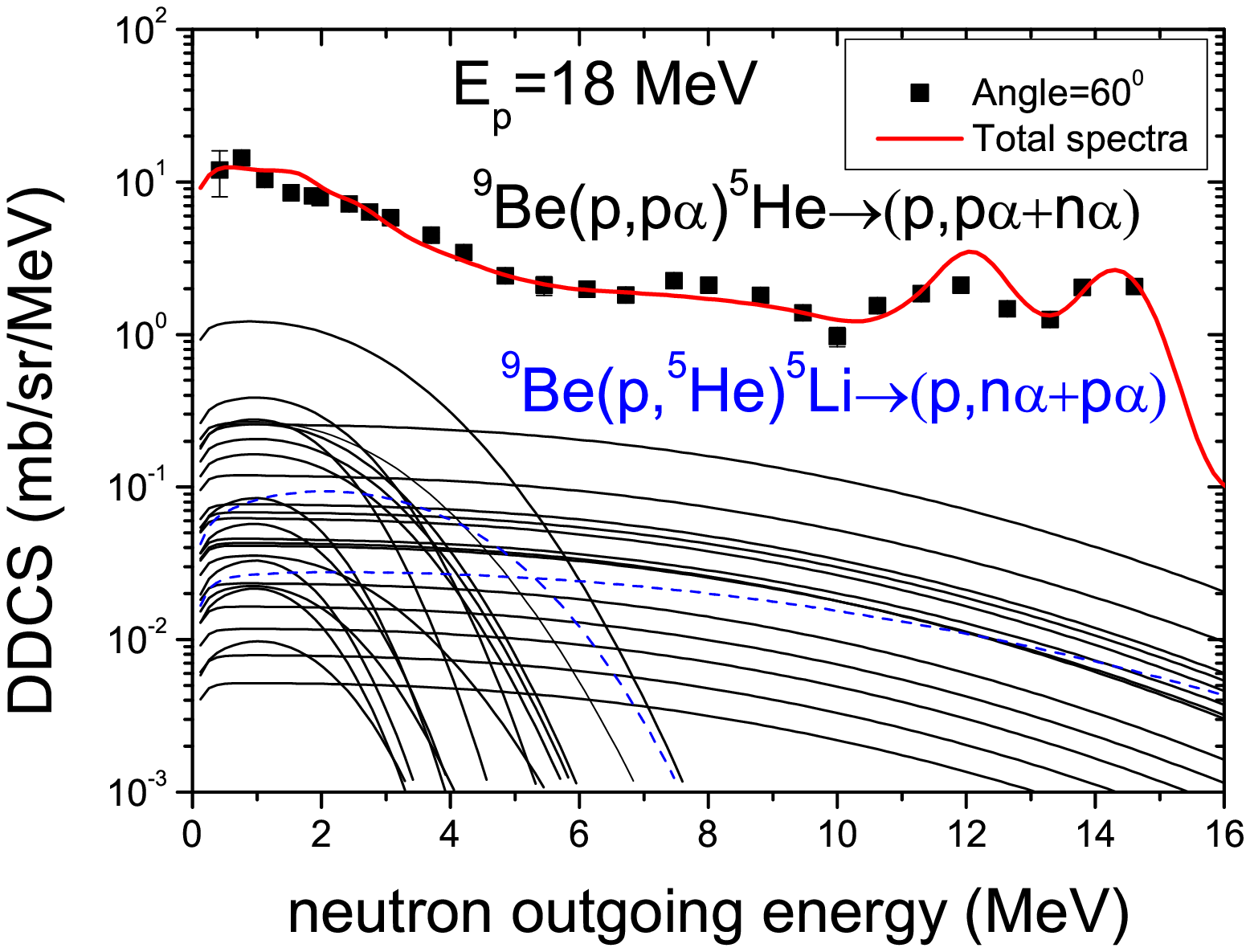}
\caption{(Color online) The same as Fig. \ref{Fig1}. But the black solid lines denote the partial spectra of the emitted neutron in reaction channel (p, p$\alpha$)$^5$He$\rightarrow$(p, p$\alpha$+n$\alpha$) from the 4th-17th excited energy levels of $^9$Be to the lowest two energy levels of the secondary residual nucleus $^5$He, which can spontaneously breakup a neutron and an alpha. And the blue dash lines denote the partial spectra of the emitted neutron in double two-body breakup channel (p, $^5$He)$^5$Li$\rightarrow$(p, n$\alpha$+p$\alpha$) from the ground state and 1st excited energy level of the $^5$He.}\label{Fig4}
\end{figure}

\section{Applications to p+$^9$Be reactions}
\label{sect3}

For neutron induced nuclear reactions with 1p-shell light nuclei involved, such as $^{6}$Li \cite{Zhang2001Li6}, $^{7}$Li \cite{Zhang2002Li7}, $^9$Be \cite{Sun2009Be9,Duan2010Be9}, $^{10}$B \cite{Zhang2003B10}, $^{11}$B \cite{Zhang2003B11}, $^{12}$C \cite{Zhang1999C12,Sun2007C12,Sun2008kerma}, $^{14}$N \cite{Yan2005N14}, $^{16}$O \cite{Zhang2001O16,Duan2005O16} and $^{19}$F \cite{Duan2007}, the calculated double-differential cross sections of outgoing neutrons agree greatly well with the experimental data. In this section, taking p+$^9$Be reaction at 18 MeV as an example, we analyze the open reaction channels in detail at 18 MeV, and calculate the double-differential cross sections of outgoing neutrons using PUNF code in the frame of STLN. The calculated results are compared with the existing experimental data, and the partial double-differential cross sections of outgoing neutron from possible energy levels are shown in detail.

\subsection{Analysis of the reaction channels}\label{sect4.1}

For proton induced $^9$Be reaction, there theoretically exists reaction channels at incident energy $E_p \leq$20 MeV in terms of the reaction threshold energies $E_{th}$ as follows
\begin{eqnarray}\label{eq4.1}
\textmd{p}+^9\textmd{Be}\rightarrow ^{10}\textmd{B}^* \rightarrow \left\{
\begin{array}{llr}
(\textmd{p}, \gamma)^{10}\textmd{B}, ~~~~Q=+6.586 \textmd{MeV}, ~~~~E_{th}=0.000 \textmd{MeV}\\
(\textmd{p}, \textmd{n})^{9}\textmd{B},~~~~Q=-1.850 \textmd{MeV}, ~~~~E_{th}=2.067 \textmd{MeV}\\
(\textmd{p}, \textmd{p})^{9}\textmd{Be},~~~~Q=~0.000 \textmd{MeV}, ~~~~E_{th}=0.000 \textmd{MeV}\\
(\textmd{p}, \alpha)^{6}\textmd{Li}, ~~~~Q=+2.127 \textmd{MeV}, ~~~~E_{th}=0.000 \textmd{MeV}\\
(\textmd{p}, ^3\textmd{He})^{7}\textmd{Li}, ~~Q=-11.202 \textmd{MeV}, ~~E_{th}=12.455 \textmd{MeV}\\
(\textmd{p}, \textmd{d})^{8}\textmd{Be}, ~~~~Q=+0.559 \textmd{MeV}, ~~~~E_{th}=0.000 \textmd{MeV}\\
(\textmd{p}, ^5\textmd{He})^{5}\textmd{Li}, ~~Q=-4.434 \textmd{MeV}, ~~~~E_{th}=4.930 \textmd{MeV}\\
(\textmd{p}, \textmd{np})^{8}\textmd{Be}, ~~Q=-1.665 \textmd{MeV}, ~~~~E_{th}=1.851 \textmd{MeV}\\
(\textmd{p}, \textmd{n}\alpha)^{5}\textmd{Li}, ~~~~Q=-3.539 \textmd{MeV}, ~~~~E_{th}=3.935 \textmd{MeV}\\
(\textmd{p}, \textmd{pn})^{8}\textmd{Be}, ~~~~Q=-1.665 \textmd{MeV}, ~~~~E_{th}=1.851 \textmd{MeV}\\
(\textmd{p}, \textmd{p}\alpha)^{5}\textmd{He}, ~~~~Q=-2.467 \textmd{MeV}, ~~~~E_{th}=2.743 \textmd{MeV}\\
(\textmd{p}, \alpha\textmd{n})^{5}\textmd{Li}, ~~~~Q=-3.539 \textmd{MeV}, ~~~~E_{th}=3.935 \textmd{MeV}\\
(\textmd{p}, \alpha\textmd{p})^{5}\textmd{He}, ~~~~Q=-2.467 \textmd{MeV}, ~~~~E_{th}=2.743 \textmd{MeV}.
\end{array}
\right.
\end{eqnarray}

Considering the conservations of the energy, angular momentum and parity in the particle emission processes, the reaction channels of the first particle emission are listed as follows
\begin{eqnarray}\label{eq4.2}
\textmd{p}+^9\textmd{Be}\rightarrow ^{10}\textmd{B}^* \rightarrow \left\{
\begin{array}{l}
\textmd{n}+ ^{9}\textmd{B}^* ~~(k_1=gs, 1, 2, ...,10)\\
\textmd{p}+ ^{9}\textmd{Be}^* ~~(k_1=gs, 1, 2, ..., 21)\\
\alpha+ ^{6}\textmd{Li}^* ~~(k_1=gs, 1, 2, ..., 7)\\
^3\textmd{He}+ ^{7}\textmd{Li}^* ~~(k_1=gs, 1, 2, ..., 4)\\
\textmd{d}+ ^8\textmd{Be}^* ~~(k_1=gs, 1, 2, ..., 8)\\
\textmd{t}+ ^7\textmd{Be} ~~(k_1=gs)\\
^5\textmd{He}+ ^5\textmd{Li}^* ~~(k_1=gs, 1).
\end{array}
\right.
\end{eqnarray}
Where, $k_1$ denotes the residual nuclei $M_1$ at $k_1$-th energy level, and $gs$ denotes its ground state.

For the first particle emission channel $^9$Be(p, n)$^9$B$^*$, the first residual nucleus $^9$B$^*$ can still emit a proton at each excited energy level, and the secondary residual nucleus $^8$Be$^*$ can spontaneously break up into two alpha \cite{Zhang2003B10}. Thus, these reaction processes belong to the reaction channel $^9$Be(p, np2$\alpha$) at the final state.

For reaction channel $^9$Be(p, p)$^9$Be$^*$, if the first residual nucleus $^9$Be$^*$ is at ground state, so this channel belongs to the compound nucleus elastic scattering. If the first residual nucleus $^9$Be$^*$ is at the $k_1$-th ($k_1\geq 1$) excited energy level, each energy level will emit a neutron and the secondary residual nucleus $^8$Be$^*$ can spontaneously break up into two alpha. Thus, these reaction processes also belong to the reaction channel $^9$Be(p, np2$\alpha$) at the final state. Especially, if the first residual nucleus $^9$Be$^*$ is at the $k_1$-th ($k_1\geq 4$) excited energy level, each energy level will be possible to emit an alpha, and the secondary residual nucleus $^5$He$^*$ can also spontaneously break up into a neutron and an alpha. Thus, these reaction processes also belong to the reaction channel $^9$Be(p, np2$\alpha$) at the final state. Therefore, the particle emission processes of the first residual nucleus $^9$Be$^*$ can be described as follows \cite{Duan2010Be9,Sun2009Be9}
\begin{eqnarray}\label{eq4.3}
^{9}\textmd{Be}^* \rightarrow \left\{
\begin{array}{l}
k=gs, ~~~~~~~~~~~~~~~~~~~~~~~~~~~~(\textmd{p}, \textmd{p})^{9}\textmd{Be}\\
k\geq 1, \textmd{n}+ ^{8}\textmd{Be}^*\rightarrow 2\alpha ~~~~~~~~~~(\textmd{p}, \textmd{np}2\alpha)\\
k\geq 4, \alpha+^{5}\textmd{He}^*\rightarrow \textmd{n}+\alpha ~~~~~~(\textmd{p}, \textmd{np}2\alpha).
\end{array}
\right.
\end{eqnarray}

For reaction channel $^9$Be(p, $\alpha$)$^6$Li$^*$, the first residual nucleus $^6$Li$^*$ is possible to emit neutron at different excited energy levels through $^6$Li$^* \rightarrow$p+$^5$He ($^5$He$\rightarrow$n+$\alpha$) and $^6$Li$^* \rightarrow$n+$^5$Li ($^5$Li$\rightarrow$p+$\alpha$) \cite{Zhang2001Li6,Zhang2002Li7}, but the cross section of this first particle emission process is so small that its contribution to neutron products can be reasonably neglected.

For reaction channel $^9$Be(p, $^3$He)$^7$Li$^*$ at $E_p=18$ MeV, the first residual nucleus $^7$Li$^*$ \cite{Zhang2001Li6,Zhang2002Li7} only at the ground, 1st and 2nd excited energy levels can not emit neutron. So the contributions of this reaction channel to neutron products are not considered in this work, as well as the reaction channels $^9$Be(p, d)$^8$Be$^*$ and $^9$Be(p, t)$^7$Be$^*$.

In addition, the double two-body breakup reaction $^9$B$^*$(p, $^5$He$^*$)$^5$Li$^*$ also belongs to channel (p, np2$\alpha$) through breakup reactions $^5$He$\rightarrow$n+$\alpha$ and $^5$Li$\rightarrow$p+$\alpha$. In conclusion, for proton induced $^9$Be reaction, there exists actually reaction channels at incident energy $E_p \leq$20 MeV as follows
\begin{eqnarray}\label{eq4.4}
\textmd{p}+^9\textmd{Be}\rightarrow ^{10}\textmd{B}^* \rightarrow \left\{
\begin{array}{lr}
\textmd{n}+^{9}\textmd{B}^* (k_1=gs, 1, ..., 10) \rightarrow \textmd{p}+^8\textmd{Be}^* \rightarrow 2\alpha, ~~~(\textmd{p, np}2\alpha)\\
\textmd{p}+^{9}\textmd{Be}^* (k_1=gs), ~~~~~~~~~~~~~~Compound~nucleus~elastic\\
~~~~~~~~~~~(k_1\geq 1) \rightarrow \textmd{n}+^8\textmd{Be}^*\rightarrow 2\alpha, ~~~~~~~~~~~~~~(\textmd{p, np}2\alpha)\\
~~~~~~~~~~~(k_1\geq 4) \rightarrow \alpha+^5\textmd{He}^*\rightarrow \textmd{n}+\alpha, ~~~~~~~~~~(\textmd{p, np}2\alpha)\\
\alpha+^{6}\textmd{Li}^* (k_1=gs, 2), ~~~~~~~~~~~~~~~~~~~~~~~~~~~~~~~~~~(\textmd{p}, \alpha)^6\textmd{Li}\\
~~~~~~~~~~~(k_1=1, 3, 4, ..., 7) \rightarrow \textmd{d}+\alpha, ~~~~~~~~~~~~~~(\textmd{p, d}2\alpha)\\
^3\textmd{He}+^7\textmd{Li}^* (k_1=gs, 1, 2, 3),~~~~~~~~~~~~~~~~~~~~~~~~~~(\textmd{p}, ^3\textmd{He})^7\textmd{Li}\\
\textmd{d}+^8\textmd{Be}^* (k_1=gs, 1, ..., 8)\rightarrow 2\alpha,~~~~~~~~~~~~~~~~~~~(\textmd{p, d} 2\alpha)\\
\textmd{t}+^7\textmd{Be} (k_1=gs),~~~~~~~~~~~~~~~~~~~~~~~~~~~~~~~~~~~~~~(\textmd{p, t})^7\textmd{Be}\\
^5\textmd{He}+^5\textmd{Li}^* (k_1=gs, 1) \rightarrow \textmd{n}+\alpha+\textmd{p}+\alpha,~~~~~~~~(\textmd{p, np}2\alpha).\\
\end{array}
\right.
\end{eqnarray}

From Eq. (\ref{eq4.4}), one can see that the contributions to the double-differential cross sections of outgoing neutron only come from reaction channel (p, np2$\alpha$), which consists of four kinds of the particle emission processes.

\subsection{Calculation of the double-differential cross sections of outgoing neutrons}

In the case of p+$^9$Be reaction at $E_p=18$ MeV with outgoing angle $60^0$ in LS, the partial double-differential cross sections of outgoing neutrons from reaction channel (p, n)$^9$B are shown in Fig. \ref{Fig1}. The black lines denote the partial neutron spectra coming from the ground state to 9th excited energy levels ($k_1=gs, 1, ..., 9$, as the numbers labeled in Fig. \ref{Fig1}) of the first residual nucleus $^9$B. Because of the level widths and energy resolution in the measurements, the measured data are always in a broadening form. Therefore, for fitting measurements the broadening effect must be taken into account in the first particle emission processes \cite{Zhang1999C12}. Only the cross sections with the values larger than 0.1 mb are given, as well as the following figures.

The partial double-differential cross sections of outgoing neutrons from reaction channel (p, pn)$^8$Be$\rightarrow$(p, pn+2$\alpha$) are shown in Fig. \ref{Fig2}, but the black lines denote the partial spectra of the secondary emitted neutron from the $k_1$-th excited energy levels ($k_1=1, 2, ..., 17$, as the numbers labeled in Fig. \ref{Fig2}) of the first residual nucleus $^9$Be to the ground state of the secondary residual nucleus $^8$Be. In Fig. \ref{Fig3}, the black solid lines denote the partial spectra of the secondary emitted neutron from the $k_1$-th excited energy levels ($k_1=6, ..., 17$, as the numbers labeled in Fig. \ref{Fig3}) of the first residual nucleus $^9$Be to the first excited energy level of the secondary residual nucleus $^8$Be. The green dash lines denote the partial spectra of the secondary emitted neutron from the $k_1$-th ($k_1=12, ..., 17$) excited energy levels of the first residual nucleus $^9$Be to the second excited energy level of the secondary residual nucleus $^8$Be. The blue dot lines denote the partial spectra of the secondary emitted neutron from the $k_1$-th ($k_1=14, ..., 17$) excited energy levels of the first residual nucleus $^9$Be to the third excited energy level of the secondary residual nucleus $^8$Be. The calculated results show that the contributions ($> 0.1$ mb) only come from two energy levels ($k_1=14$ and 17) of $^9$Be.

In Fig. \ref{Fig4}, the black solid lines denote the partial spectra of the emitted neutron from reaction channel (p, p$\alpha$)$^5$He$\rightarrow$(p, p$\alpha$+n$\alpha$). The contributions of these partial neutron spectra come from the emissions between the 4th-17th excited energy levels of the first residual nucleus $^9$Be and the lowest two energy levels of the secondary residual nucleus $^5$He, which can spontaneously breakup a neutron and an alpha. And the blue dash lines denote the partial spectra of the emitted neutron from reaction channel (p, $^5$He)$^5$Li$\rightarrow$(p, n$\alpha$+p$\alpha$). The contributions of these partial neutron spectra come from the ground state and the 1st excited energy level of $^5$He.

\begin{figure}
\centering
\includegraphics[width=24cm,angle=0]{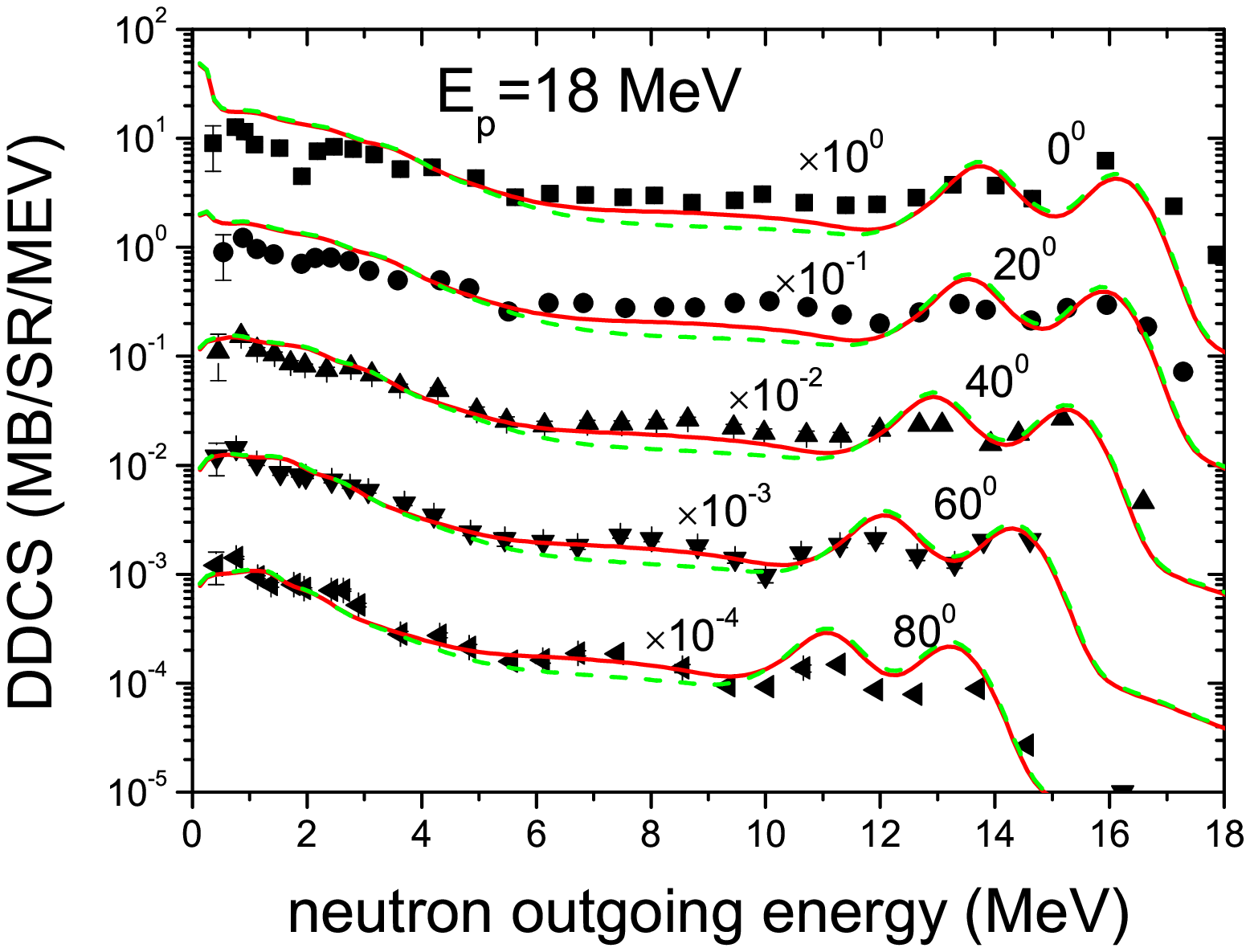}
\caption{(Color online) The total double-differential cross sections of outgoing neutron for reaction $^9$Be(p, xn) with outgoing angles 0$^0$, 20$^0$, 40$^0$, 60$^0$ and 80$^0$ at $E_p$=18 MeV in LS, respectively. The points denote to the experimental data taken from Ref. \cite{Verbinski1969}. The red solid lines denote the calculated results using the real and two predicted energy levels of $^9$Be, and the green dash lines denote the calculated results only using the real energy levels of $^9$Be.}\label{Fig5}
\end{figure}

\begin{figure}
\centering
\includegraphics[width=24cm,angle=0]{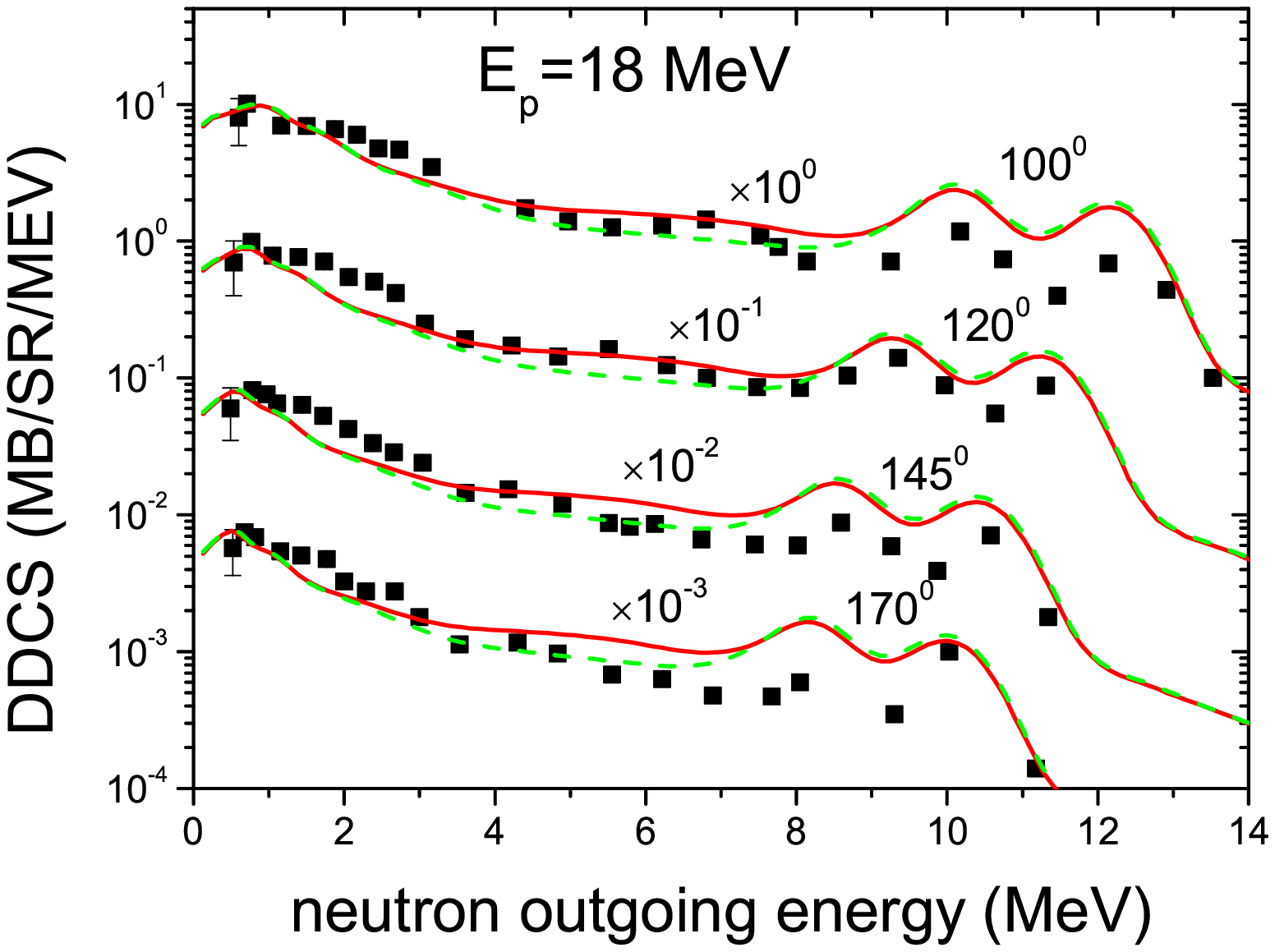}
\caption{(Color online) The same as Fig. \ref{Fig5}, but with outgoing angles 100$^0$, 120$^0$, 145$^0$ and 170$^0$, respectively. }\label{Fig6}
\end{figure}

Summing up all of the partial double-differential cross sections of outgoing neutrons, we can obtain the total double-differential cross sections at $E_p=18$ MeV with outgoing angle $60^0$ (as shown the red lines in Figs. \ref{Fig1}-\ref{Fig4}). In these figures, the points denote the experimental data measured by Verbinski et al \cite{Verbinski1969}. One can see that the calculated total double-differential cross sections of outgoing neutrons agree greatly well with the experimental data. Similarly, the calculated total double-differential cross sections of outgoing neutrons at other angles also agree well with the experimental data as shown in Fig. \ref{Fig5} and Fig. \ref{Fig6}.

It is worth mentioning that all of the final states is the discrete levels of the residual nuclei in light nucleus reactions. Therefore, the theoretical calculations are very sensitive to the level schemes of the target and residual nuclei. Although the updated energy level schemes of the target and the residual nuclei \cite{Tilley2002,Tilley2004} are employed for reaction $^9$Be(p, xn), the contributions coming from the real 9th and 10th energy levels of the target nucleus $^9$Be, as shown the green dash lines in Fig. \ref{Fig5} and Fig. \ref{Fig6}, are still deficiencies. So two predicted levels 9.0($\frac{5}{2}^+$) and 10.0($\frac{5}{2}^+$) between the 9th and 10th levels have been employed in this paper as neutron induced $^9$Be reaction \cite{Sun2009Be9}. In Fig. \ref{Fig5} and Fig. \ref{Fig6}, the red solid lines denote the results using the real energy levels and two predicted energy levels of $^9$Be, and the green dash lines denote the results only using the real energy levels, respectively. One can see that the calculated results of adding two predicted levels agree better with the existing experimental data.

\section{Summary}
\label{sect4}

Our previous studies indicate that the calculated double-differential cross sections agree greatly well with the experimental data for neutron induced nuclear reactions with 1p-shell light nuclei involved, which have been successfully used to set up file-6 in CENDL 3.1 library. In this paper, STLN is proposed to describe the light particle induced nuclear reactions with 1p-shell light nuclei involved. In the dynamic of STLN, not only the angular momentum and parity conservations for both equilibrium and pre-equilibrium processes, but also the Coulomb barriers of the incident and outgoing charged particles are considered in different particle emission processes. In the kinematics of STLN, the recoiling effects in various emission processes are taken strictly into account. Taking $^9$Be(p, xn) reaction as an example, we further calculate the double-differential cross sections of outgoing neutrons and charged particles using PUNF code in the frame of STLN. The calculated results agree very well with the existing experimental neutron double-differential cross sections at $E_p=18$ MeV, and indicate that PUNF code is a powerful tool to set up file-6 in the reaction data library for the light particle induced nuclear reactions with 1p-shell light nuclei involved. However, it is worth mentioning that STLN and PUNF code are applied to nuclear reactions without polarization of incoming light particles and orientation of target nuclei.

In addition, two predicted levels of target nucleus $^9$Be have been employed in this paper. The great agreement between the calculated results and the experimental data shows again that there may be lack of several levels of $^9$Be. We hope these predicted levels could be validated by experiment in the future.
For analytically describing the double-differential cross sections of reaction products in the sequential particle emission processes, a new integral formula, which has not been compiled in any integral tables or mathematical softwares, is employed to obtain analytical Legendre expansion coefficients. This integral formula can largely reduce the volume of file-6 in nuclear reaction databases with full energy balance. This integral formula and STLN are being tested by light charged particles induced nuclear reactions with 1p-shell light nuclei involved.

\textbf{Acknowledgements}

We wish to thank Dr. Y. Guo, professors S. G. Zhou, N. Wang, L. Ou and M. Liu for some valuable suggestions. This work is supported by the National Natural Science Foundation of China (No. 11465005); the Natural Science Foundation of Guangxi (No. 2014GXNSFDA118003); Guangxi University Science and Technology Research Project (No. 2013ZD007); the Open Project Program of State Key Laboratory of Theoretical Physics, Institute of Theoretical Physics, Chinese Academy of Sciences, China (No. Y4KF041CJ1);
and the project of outstanding young teachers' training in higher education institutions of Guangxi.

\end{document}